\documentclass[iop]{emulateapj}

\usepackage{multirow}
\usepackage{placeins}
\usepackage{paralist}
\usepackage{graphics,graphicx}

\begin{document}

\title{Common Envelope Mechanisms: Constraints from the X-ray Luminosity Function of High Mass X-ray Binaries}
\author{Zhao-Yu Zuo$^{1,3}$ and Xiang-Dong Li$^{2,3}$}
\affil{$^1$School of Science, Xi'an Jiaotong University, Xi'an 710049, China\\
$^2$Department of Astronomy, Nanjing University, Nanjing 210093, China\\
$^3$Key laboratory of Modern Astronomy and Astrophysics (Nanjing
University), Ministry of Education, Nanjing 210093, China
\\zuozyu@mail.xjtu.edu.cn; lixd@nju.edu.cn}

\begin{abstract}

We use the measured X-ray luminosity function (XLF) of high-mass X-ray binaries (HMXBs) in nearby star-forming galaxies to constrain the common envelope (CE) mechanisms, which play a key role in governing the binary evolution. We find that the XLF can be reproduced quite closely under both CE mechanisms usually adopted, i.e., the $\alpha_{\rm CE}$ formalism and the $\gamma$ algorithm, with a reasonable range of parameters considered. Provided that the parameter combination is the same, the $\gamma$ algorithm is likely to produce more HMXBs than the $\alpha_{\rm CE}$ formalism, by a factor of up to $\sim$ 10. In the framework of the $\alpha_{\rm CE}$ formalism, a high value of $\alpha_{\rm CE}$ is required to fit the observed XLF, though it does not significantly affect the global number of the HMXB populations. We present the detailed components of the HMXB populations under the $\gamma$ algorithm and compare them with those in Zuo et al. and observations. We suggest the distinct observational properties, as well as period distributions of HMXBs, may provide further clues to discriminate between these two types of CE mechanisms.

\end{abstract}


\section{Introduction}

The common envelope (CE) evolution is among the most important and least well-constrained processes in binary evolution. It is commonly thought to occur if the mass transfer is dynamically unstable. The result is that the accreting star spirals into the envelope of the donor star \citep[see][for reviews]{iben93,taam00,webbink08,taam10,Ivanova13}. The orbital energy and angular momentum of the accreting star are then transferred into the CE via an as of yet unknown mechanism. This may end with a stellar merger or, if the binary can survive, a binary with a much shorter orbital period. The CE evolution is critical in the formation of various kinds of compact binaries.

There have been extensive three dimensional hydrodynamical simulations \citep[e.g.,][]{rl96,stc98,stb00,Fryxell00,OShea05,Fryer06,Passy12,rt12}. However the physics of CE evolution still remains poorly understood, primarily due to a mix of various kinds of physical processes operating over a large range of timescales and length scales during the CE phase. Due to the difficulties in modeling the detailed CE evolution, population synthesis simulations commonly resort to simplified and parameterized descriptions to relate the post- and pre-CE orbital parameters \citep{ty79}. One such parametrization dictates the CE phase in terms of a simple energy budget \citep[known as the $\alpha_{\rm CE}$ formalism,][]{heuvel76,Webbink84,ls88,iben93} and the other in terms of the angular momentum budget \citep[named as the $\gamma$ algorithm,][]{nvyp00,nt05}. Both approaches have the power to account for some specific classes of post-CE binaries (PCEBs), such as cataclysmic variables \citep[CVs,][]{king88}, subdwarf B binaries \citep{mr03,han02,han03}, low-mass X-ray binaries \citep{cc06}, and other compact objects thought to have suffered a merger, which are probably responsible for gamma-ray bursts \citep{fryer99,Thone} and Type Ia supernova \citep[SNe Ia,][]{it84,bbr05,ruiter11,meng11,my12,wh12}. There is an energetic debate over the two approaches in the literature.

High-mass X-ray binaries (HMXBs) are good examples of the effect of CE evolution. Luminous HMXBs usually have experienced CE evolution so that they have close orbits, which lead to a high wind-capture rate by the compact star. Tight orbits also help the binary survive the SN kick during the formation of the compact star. However, if the initial binary orbit is not large enough, CE evolution may lead to mergers, reducing the HMXB production. Thus, the populations and the specific characteristics (for example the orbital period distribution) of HMXBs can be used to probe the physical interactions during the CE phase.
 
The formation of HMXBs involves several evolutionary pathways \citep[][see also Tauris \& van den Heuvel 2006]{bv00,Linden10}. Beginning with two relatively massive stars ($\gtrsim\,10\,M_{\odot}$), the more massive primary evolves and commences mass transfer to the secondary. The mass transfer can be either dynamically stable or unstable. In the latter case, CE evolution occurs that greatly shrinks the binary orbit. The resultant binary consists of the primary's core and the secondary. The primary's core then collapses to form a neutron star (NS) or black hole (BH). An HMXB appears when the compact star is able to accrete from the secondary by capture of the stellar wind or Roche lobe overflow (RLOF). Note that the secondary can be on the main-sequence (MS) or a (super)giant star. In some cases the second mass transfer may also lead to a CE phase, during which the envelope of the secondary is stripped, leaving a naked helium core. This leads to the formation of HMXBs with Wolf-Rayet companions.
 
HMXBs have some unique statistical characteristics \citep[for catalogs, see][]{lph05,lph06}. One of the most striking 
 features is that their X-ray luminosity function (XLF) follows a universal power law form over a broad X-ray luminosity range, from $\sim 10^{35}$ to $\sim 10^{40} \rm \,erg\,s^{-1}$. This was first discovered by \citet{ggs03}, and further confirmed 
 recently by \citet[][hereafter MGS12 for short]{mineo12}. The XLF has been shown to follow a power law with a single slope of $\sim$ 1.6, without any significant feature near the critical Eddington luminosity of an NS or a stellar mass BH. Additionally, the collective luminosity of HMXB populations scales with the star formation rate (SFR) as $L_{\rm X} (\rm erg\,s^{-1}) \approx 2.6\cdot10^{39}\times \rm SFR (M_{\odot}\,yr^{-1})$.

In the present work, we apply the updated evolutionary population synthesis (EPS) techniques to model the XLF of HMXBs, taking into account both the $\alpha_{\rm CE}$ algorithm and $\gamma$ algorithm (with different choices of $\alpha_{\rm CE}$ and $\gamma$, respectively), to describe the CE evolution. By comparing the observational sample with our theoretical expectations, we try to discriminate or constrain the effects of the two CE mechanisms.

This paper is organized as follows. In \S 2 we describe the
population synthesis method and the input physics for X-ray binaries (XRBs) in our
model. The calculated results and discussions are presented in \S 3. Our conclusions 
are in \S 4.

\section{MODEL DESCRIPTION}

We use the EPS code developed by \citet{Hurley00,Hurley02} and recently updated by \citet{zuo14a} to calculate the expected number and the X-ray luminosity of HMXBs. In the present code, the model for compact object formation has been significantly revised by taking into account the formation of NSs through electron capture supernovae \citep[ECS,][]{Podsiadlowski04} and the fallback process for both delayed and direct BH formation during core collapse \citep{fk01}. The prescriptions for the wind mass loss rates of massive stars \citep[][see also Belczynski et al. 2010]{vink01} and the compact remnant masses \citep[][see also Belczynski et al. 2012]{fryer12} are adopted in the code. We also update the criteria for CE occurrence as described below.

\subsection{The CE Phase}
During the binary evolution, the mass ratio ($q=M_{\rm donor}/M_{\rm accretor}$) is a crucial factor determining the stability of mass transfer. If it is larger than a critical value, $q_{\rm crit}$, the mass transfer is dynamically unstable and a CE phase follows \citep{p76}. The ratio $q_{\rm crit}$ varies with the evolutionary state of the donor star at the onset of RLOF and the mass loss mechanisms during the mass transfer \citep{hw87,w88,prp02,ch08}. In this study, we adopt an updated $q_{\rm crit}$ for Hertzsprung gap donor star, recently calculated by \citet[][also see the Appendix A in Zuo, Li \& Gu 2014a for more details]{shao12}. If the primary is on the first giant branch (FGB) or the asymptotic giant branch (AGB), we use
\begin{equation}
q_{\rm crit}=[1.67-x+2(\frac{M_{\rm c1}}{M_1})^5]/2.13
\end{equation}
where $M_{\rm c1}$ is the core mass of the donor star, and $x$=d ln$R_1/$d ln$M$
is the mass-radius exponent of the donor star. If the mass donor star is
a naked helium giant, $q_{\rm crit}=0.784$ \citep[see][for more details]{Hurley02}.

\subsubsection{The $\alpha_{\rm CE}$ formalism}

In the energy budget approach, the CE evolution is parameterized in terms of
the orbital energy and binding energy as $E_{\rm bind} \equiv \alpha_{\rm CE}
\triangle E_{\rm orb}$ \citep{Webbink84,webbink08}, where the parameter $\alpha_{\rm CE}$ describes the efficiency of converting the orbital energy (\textbf{$E_{\rm orb}$}) into the kinetic energy, which is used to eject the envelope, and $E_{\rm bind}$ is the binding energy of the envelope. The CE evolution is governed by the following equation \citep{Kiel06}:
\begin{equation}
\alpha_{\rm CE}[\frac{GM_{\rm c}M_{2}}{2 a_{\rm f}}-\frac{GM_{\rm
c}M_{2}}{2 a_{\rm i}}]=-\frac{GM_1M_{\rm env}}{R_{L_1}\lambda},
\end{equation}
which yields the ratio of final (post-CE) and initial (pre-CE) orbital
separations as
\begin{equation}
\frac{a_{\rm f}}{a_{\rm i}}= \frac{M_{\rm c}M_{2}}{M_{1}}
\frac{1}{M_{\rm c}M_2/M_1+2M_{\rm env}/(\alpha_{\rm CE} \lambda
R_{\rm L1})},
\end{equation}
where $G$ is the gravitational constant, $M_{\rm c}$ the helium-core mass of the
primary star (of mass $M_1$), $M_2$ the mass of the secondary star, $R_{L_1}$ the RL radius of the primary star, $M_{\rm env}$ 
the mass of the primary's envelope, $a_i$ and $a_f$ denote the initial and final orbital
separations, respectively, and $\lambda$ is a parameter related to the stellar mass-density distribution.

The $\lambda$ value depends on the structure and evolution of the donor star. However, in previous studies, it was usually adopted as constant ($\sim 0.5$) for simplicity \citep{Hurley02,zuo08}. Here we calculate the values of $\lambda$ from detailed stellar models including the contribution from the internal (and ionization) energies within the envelope \citep[][also see Xu \& Li 2010 and Loveridge et al. 2011]{zuo14a}.

We consider three constant, global values of $\alpha_{\rm CE}$. For our basic model,
we use $\alpha_{\rm CE}=0.5$ \citep{zuo14a}. We also consider two other extreme
values of $\alpha_{\rm CE}=1.0$ and 0.1 since $\alpha_{\rm CE}$ is expected to be no more than unity if we consider the internal energies in calculating $E_{\rm bind}$. 
Different CE efficiencies for the first and second CE episodes are also examined to test its effect on the XLFs. Models with different values of $\alpha_{\rm CE}$ are denoted as A01A01, A05A05, A10A10, A01A05, and A05A01, respectively, where the two digits following each letter correspond to the values of $\alpha_{\rm CE}$ 
during the first and second CE episodes, respectively.

Alternatively, recent studies on WD binaries show that $\alpha_{\rm CE}$ may be a function of binary parameters rather than constant \citep{pw07,zgn00,marco11,dkk12}, although the final relationship has not yet been well developed.
Following \citet{marco11}, we adopt
\begin{equation}
\alpha_{\rm CE}=0.05\times q^{1.2},
\end{equation}
where $q$ is the ratio of the donor's mass to the accretor's mass at the
time of the CE interaction, and this model is denoted as AqAq.

\subsubsection{The $\gamma$ algorithm}

In the angular momentum budget approach, the CE interaction is parameterized in terms of
$\gamma$, the ratio of the fraction of angular momentum lost, and the fraction of mass loss:
\begin{equation}
\frac{\triangle J}{J}=\gamma \frac{M_{\rm env}}{M_1+M_2}
\end{equation}
where $\triangle J$ is the change of the total angular momentum ($J$) during the CE phase.
Implicitly assuming the conservation of energy, the orbital separation after the CE
is then given by
\begin{equation}
\frac{a_{\rm f}}{a_{\rm i}}=(\frac{M_1}{M_{\rm c}})^2(\frac{M_{\rm c}+M_2}{M_1+M_2})
[1-\gamma (\frac{M_{\rm env}}{M_1+M_2})]^2
\end{equation}

This description was first suggested by \citet{nvyp00} in their investigation
of the formation of double WD binaries. They found that when the energy approach is
applied to describe the first CE phase, a negative value of $\alpha_{\rm CE}$ is required,
which is clearly unphysical. Among the possible solutions leading to the known close 
double WDs, \citet{nt05} found that $0.5<\gamma<3$ for the first (putative) 
CE phase, and $1<\gamma<4$ for the second CE phase. They noted that 
a value of $\gamma$ between 1.5 and 1.75 can account for all known observed PCEBs, including double WDs, pre-CVs, and sdB plus MS binaries.

For the $\gamma$ algorithm, we consider several constant, global values 
of $\gamma$ from 1.7 to 1.0, as well as different $\gamma$ values for the first and second CE 
episodes in our calculation. These models are denoted as G17G17, G15G15, 
G13G13, G10G10, G10G17 and G17G10 where the two digits following each 
letter correspond to the values of $\gamma$ during the first and second CE 
episodes, respectively.

In the study, we first compare the two mechanisms under the same assumptions, as 
described below. The parameter combination is kept the same as in \citet{zuo14a}, where 
the best-fit model in the $\alpha_{\rm CE}$ formalism is achieved. In this case, only values 
of $\gamma$ and $\alpha_{\rm CE}$ are changed to see 
their effects on the XLF. Then we manage to determine the best-fit model in the $\gamma$ algorithm by 
varying all the key parameters, and see their effects on the XLF (see Table~1). 
Finally, the two mechanisms are compared under each best-fit model 
(i.e., model A05A05 vs. model M1).

\begin{table}
\caption{Parameters adopted for each model under the $\gamma$ algorithm.
Here $q_0$ is the initial mass ratio, IMF is the initial mass function, $f$ binary fraction, 
$\eta_{\rm Edd, BH}$ - the factor of super-Eddington accretion rate allowed for 
BH XRBs, $\sigma_{\rm kick}$ is the dispersion of kick velocity, $\eta_{\rm bol, BH(NS)}$ 
is the bolometric correction factor for BH(NS) XRBs, STD is the standard stellar winds 
while WEAK represents the standard wind mass loss rate reduced to 50\%, MT87 
represents the IMF of \citet{mt87}. In the best-fit model of $\gamma$ algorithm (M1), the parameters 
are as follows: SFH=50 Myr, $\alpha=0$, $\eta_{\rm Edd, BH}$=20, $f=0.5$, 
$\sigma_{\rm kick} = 110 \,\rm km\,s^{-1}$, $\eta_{\rm bol, BH}$=0.2,  
$\eta_{\rm bol, NS}$=0.1 and Salpeter IMF.} \centering
\label{tab:m7}
\begin{tabular}{cccccccc}\hline\hline
   Model & $P(q_0)$  & IMF &  $f$  &  $\eta_{\rm Edd, BH}$   & $\sigma_{\rm kick}$  & winds\\ \hline
      M1   & $\propto q_0^{0}$ & Salpeter     &   0.5 & 20  & 110 & STD  \\
      M2   & $\propto q_0^{0}$ & Salpeter     &   0.5 & 20  & 110 & WEAK  \\       
      M3   & $\propto q_0^{0}$ & Salpeter     &   0.8 & 20  & 110 & STD  \\
      M4   & $\propto q_0^{0}$ & Salpeter     &   0.5 & 100 & 110 & STD  \\
      M5   & $\propto q_0^{1}$ & Salpeter     &   0.5 & 20  & 110 & STD  \\
      M6   & $\propto q_0^{0}$ & MT87           &   0.5 & 20  & 110 & STD  \\
      M7   & $\propto q_0^{0}$ & Salpeter     &   0.5 & 20  & 190  & STD  \\
      M8   & $\propto q_0^{0}$ & Salpeter     &   0.5 & 20  & 265 & STD  \\  \hline
\end{tabular}
\end{table}


\subsection{Input Parameters}

We follow the evolution of a large number of binary systems, initially consisting of two
zero-age MS stars. As HMXBs in the MGS12 samples reside in nearby
star-forming galaxies, we adopt a constant SFR 
for 50 Myr and a fixed subsolar metallicity ($0.5\,Z_{\odot}$, where $Z_{\odot}=0.02$) accordingly 
\citep[see][for details]{zuo14a}. Since the observed average XLF 
has already been normalized, we choose a Salpeter initial mass function (IMF) and set the mass 
range as $0.1-100\,\rm M_{\odot}$ for the normalization in order to be in 
parallel with MGS12\footnote{We have improperly chosen an IMF of 
\citet{Kroupa01} and set the mass range as $5.0\,M_{\odot} - +\infty$ 
for the normalization in \citet{zuo14a}. The predicted HMXBs were 
overestimated by a factor of $\sim 3$, but the results and basic conclusions 
of the paper remain largely unchanged \citep[see][for details]{zuo14d}.}.
We evolve $10^6$ primordial
systems\footnote{We also change the number of the binary systems
by a factor of eight, and find no significant difference in the
final results.} and set up the same grid of initial parameters (primary mass, secondary
mass and orbital separation) as in \citet{Hurley02}.

For the initial secondary's mass ($M_2$), a power law distribution of
$P(q_0)\propto q_0^{\alpha}$ is assumed, where $q_0\equiv M_2/M_1$. In our basic model,
a flat distribution is assumed, i.e., $\alpha=0$. We adopt a
logarithmically flat distribution of initial orbital separations $\ln a$ \citep{Hurley02}.

We assume a binary fraction $f=0.5$ and that all binaries are initially in a
circular orbit. For the SN kicks imparted on an NS, we assume a Maxwellian distribution with $\sigma_{\rm kick}=
110\, \rm km\,s^{-1}$ \citep{zuo14d}. For compact objects formed with partial
mass fallback, the natal kicks are decreased by a factor of (1-$f_{\rm b}$)
where $f_{\rm b}$ is the fraction of the stellar envelope that falls back
after the SN explosion.

\begin{figure*}
  \centering
   \includegraphics[width=6in]{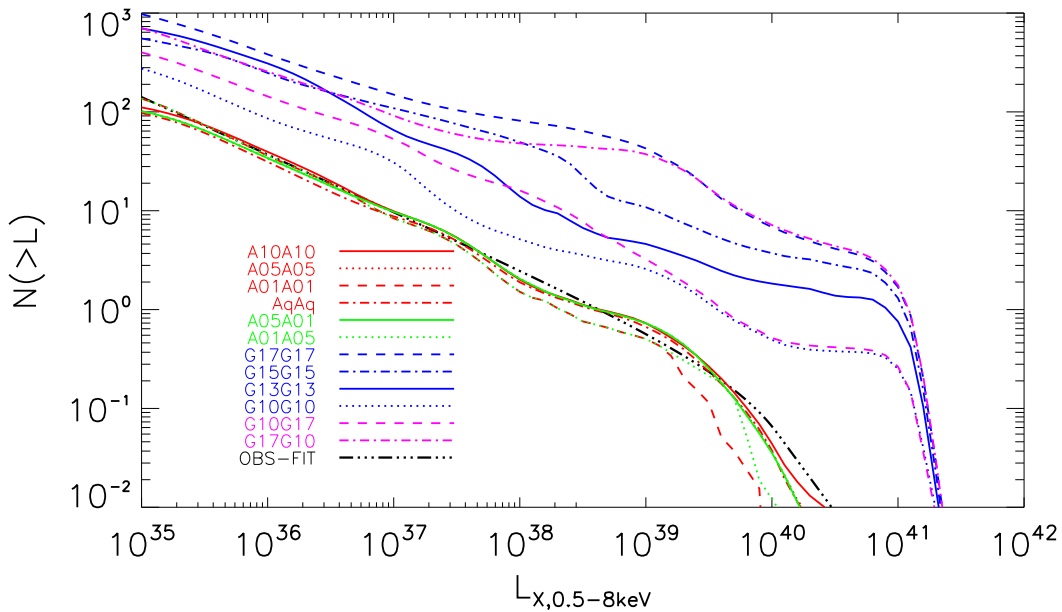}
\caption{Simulated XLFs of different models on the treatment of
the CE phase. The thick triple dot-dashed line represents the observed average XLF
(labeled as `OBS-FIT') derived by MGS12 using the data of 29 nearby star-forming
galaxies.}
  \label{Fig. 1a}
\end{figure*}

\subsection{X-ray luminosity and source type}
We adopt the same procedures to compute the $0.5-8$ keV X-ray
luminosity for MS/super-giant (SG) HMXBs and Be-XRBs  as in \citet{zuo14a}. For wind accretion, we use the classical \citet{Bondi44} formula to calculate the mass transfer rate to
the compact star. In the case of RLOF, we discriminate transient
and persistent sources using the criteria in \citet[][i.e., Eq.~36 therein]{l01} for MS
and red giant (RG) donor. The corresponding X-ray luminosity is calculated as follows:
\begin{eqnarray}
&&L_{\rm X, 0.5-8 keV}\nonumber\\
&&=\left\{
\begin{array} { ll}
  \eta_{\rm bol}\eta_{\rm out}L_{\rm Edd}&\ \rm transients\ in\ outbursts, \\
  \eta_{\rm bol}\min(L_{\rm bol},\eta_{\rm Edd}L_{\rm Edd})&\ \rm persistent\
  systems,
\end{array}
\right.
\end{eqnarray}
where $\eta_{\rm bol}$ is the bolometric correction factor
converting the bolometric luminosity ($L_{\rm bol}$) to the $0.5-8$ keV
X-ray luminosity, ranging between $\sim 0.1$ and $\sim 0.8$ 
\citep{bel08}; $L_{\rm bol} \simeq 0.1\dot{M}_{\rm acc}c^2$ where 
$\dot{M}_{\rm acc}$ is the average mass accretion rate and $c$ is 
the velocity of light. The critical Eddington luminosity $L_{\rm Edd} \simeq 4\pi GMm_{\rm
p}c/\sigma_{T}=1.3 \times 10^{38}m$\,erg\,s$^{-1}$ (where
$\sigma_{T}$ is the Thomson cross section, $m_{\rm p}$ the proton
mass, and $m$ the accretor mass in the units of solar mass). We introduce the `Begelman' factor 
$\eta_{\rm Edd}$ to allow super-Eddington luminosities.
We fix $\eta_{\rm Edd, NS}=5$ for NS XRBs \citep{zuo14a}; for BH XRBs, 
$\eta_{\rm Edd, BH}$ is set as a free parameter in the study. 
For transient sources, the outburst luminosity is taken as a fraction ($\eta_{\rm out}$) of the critical
Eddington luminosity. We take $\eta_{\rm out}=0.1$ and 1 for NS(BH) transients with
orbital period $P_{\rm orb}$ less and longer than 1 day (10 hr), respectively \citep{chen97,Garcia03,bel08}.

For Be-XRBs we employ a phenomenological definition as in 
\citet[][also see Belczynski \& Ziolkowski, 2009]{zuo14a}. 
Technically, we randomly select 25\% \citep[$f_{\rm Be}=0.25$,][]{s88,z02,mg05} of NS
binaries hosting a ($3.0\,M_{\odot}-20.0\,M_{\odot}$) B/O star to be Be-XRBs, and estimate their 
numbers. The X-ray luminosity of a Be-XRB is calculated using the empirical relation (Eq.~11) in 
\citet{dll06}, which is based on the data compiled by \citet{rp05}. Considering the duration of type I outbursts in Be-XRBs \citep[$\sim 0.2-0.3 P_{\rm orb}$,][]{reig11}, 
we adopt an upper value of the duty cycle $DC_{\rm max}=0.3$ to calculate the source numbers.

\section{Results}

\begin{figure*}
  \centering
   \includegraphics[width=6in]{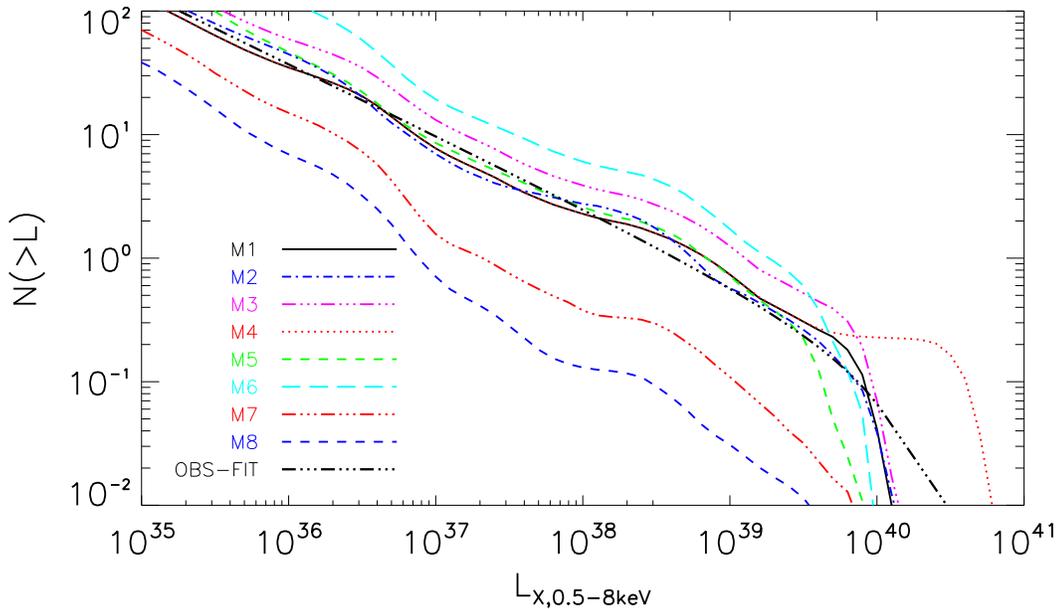}
\caption{Comparisons of the simulated (models M1-M8, all under the $\gamma-$algorithm) 
and observed average (labeled as `OBS-FIT', thick triple dot-dashed line) XLFs.
Compared to the basic model M1 (solid line), in M2 (dash-dotted line),
the standard wind mass loss rate is reduced by a factor of 2. In M3 (triple dot-dashed line), 
the binary fraction $f$ is set as 0.8. In M4 (dotted line), the factor for super-Eddington accretion 
rate for BHs is set as 100 for comparison. We take an atypical distributions of mass ratio in M5 (short-dashed
line) and a flatter IMF in M6 (long-dashed line), respectively. In models M7 and M8, the dispersion 
of kick velocity is increased to $\sigma_{\rm kick} = 190 \,\rm km\,s^{-1}$ (triple dot-dashed line) and 
$\sigma_{\rm kick} = 265 \,\rm km\,s^{-1}$ (short-dashed line), respectively.}
  \label{Fig. 1a}
\end{figure*}

\begin{figure*}
  \centering
  \includegraphics[width=6in]{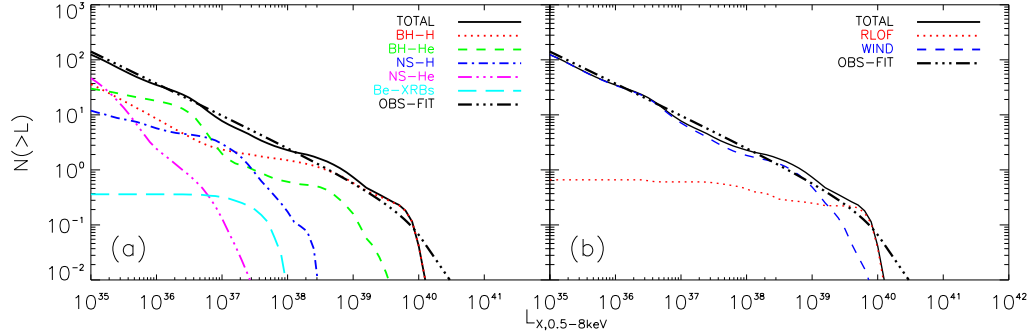}
\caption{ Detailed components of the simulated XLF (panel a)
and the accretion modes of simulated XRBs (panel b) in model M1. Panel (a): The solid,
dotted, short-dashed, dash-dotted, triple dot-dashed, long-dashed lines represent ALL-XRBs,
BH-H, BH-He, NS-H, NS-He MS/SGXRBs and Be-XRBs, respectively. Panel (b): The solid,
dotted, dashed lines represent ALL-XRBs, RLOF-fed XRBs and wind-fed XRBs, respectively.
The thick triple dot-dashed line represents the derived average XLF (labeled as
`OBS-FIT') by MGS12 using the data of 29 nearby star-forming galaxies. }
  \label{Fig. 1a}
\end{figure*}


We first compare the results in the $\alpha_{\rm CE}$ formalism and  
$\gamma$ algorithm under the same input parameters: SFH=50 Myr, $\alpha=0$, $\eta_{\rm Edd, BH}$=100, 
$f=0.5$, $\sigma_{\rm kick} = 110 \,\rm km\,s^{-1}$, $\eta_{\rm bol, BH}$=0.6,  
$\eta_{\rm bol, NS}$=0.3 and Salpeter IMF \citep{zuo14a}.
For each CE episode, models are designed by changing only one parameter 
each time to test its effect. Figure~1 compares the simulated XLFs with different treatments 
of the CE phase. Clearly, under the same parameter combination the $\gamma$ algorithm 
can produce more (up to one order of magnitude) HMXBs than the $\alpha_{\rm CE}$ formalism. 
In the framework of the $\alpha_{\rm CE}$ formalism, though all models can fit the 
observed XLF quite closely in most of the luminosity range (i.e., $10^{35}-\sim 10^{39}\,\rm 
ergs\,s^{-1}$), a high value of $\alpha_{\rm CE}$ seems more preferable. This is mainly due 
to the sparseness of short period RLOF HMXBs in the case of smaller $\alpha_{\rm CE}$ 
\citep[compare with the right panel of Figure~1 in][]{zuo14a}, the progenitors of which coalesce during the binary evolution, 
especially in the first CE phase (see models A05A01 and A01A01). In the case of $\gamma$ algorithm, 
the normalization of the simulated XLFs is rather sensitive to the value of $\gamma$, especially 
in the first CE phase (see models G10G17 and G17G17 or models G10G10 and G17G10). 
Smaller values of $\gamma$ give a better fit to the observed XLF, not only in the normalization, 
but also in the overall shape. Considering that many parameters may considerably influence the XLF \citep{zuo14a}, further thorough parameter studies are needed to determine the best-fit model in the $\gamma$ algorithm.


\begin{figure*}
  \centering
      \includegraphics[width=3in]{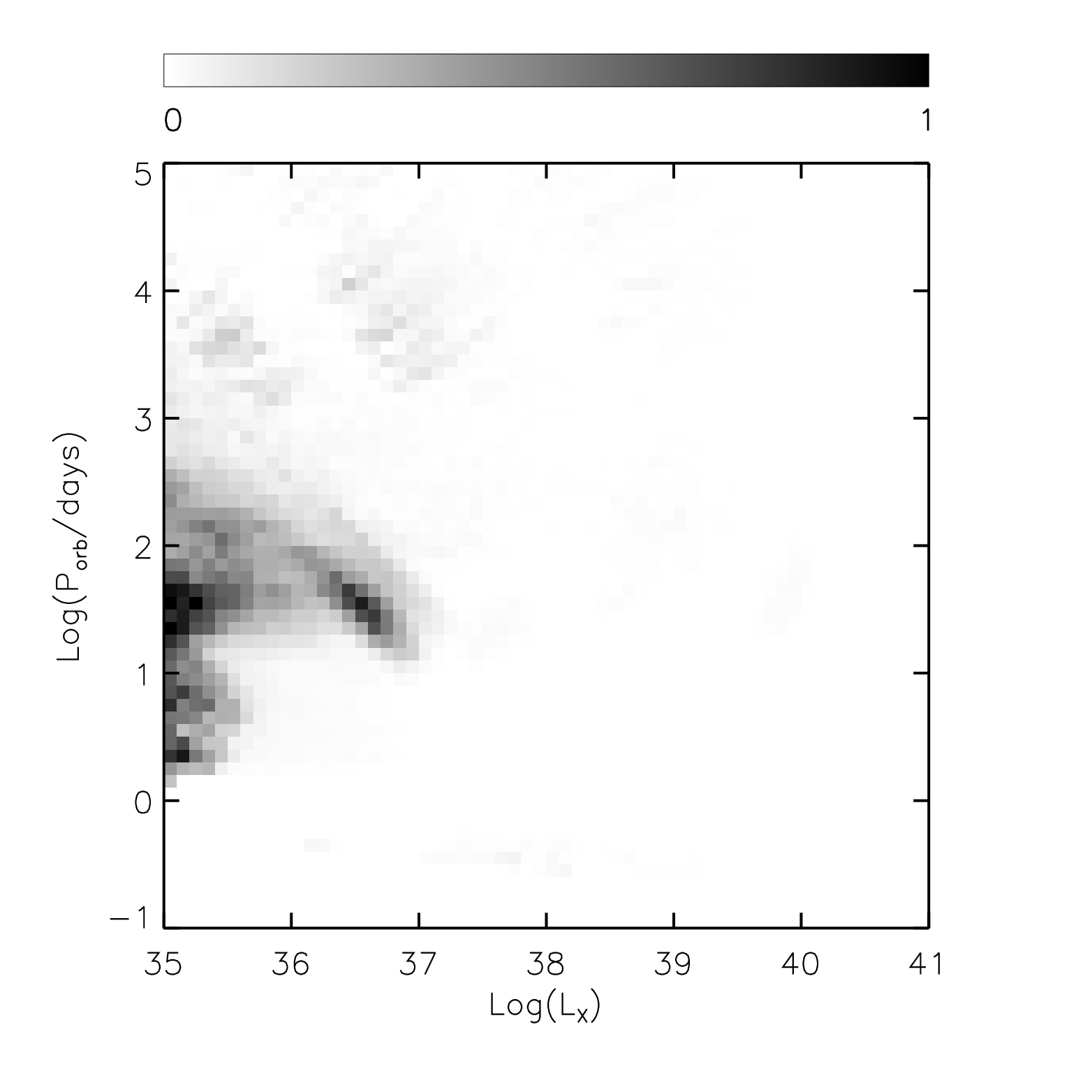}
      \includegraphics[width=3in]{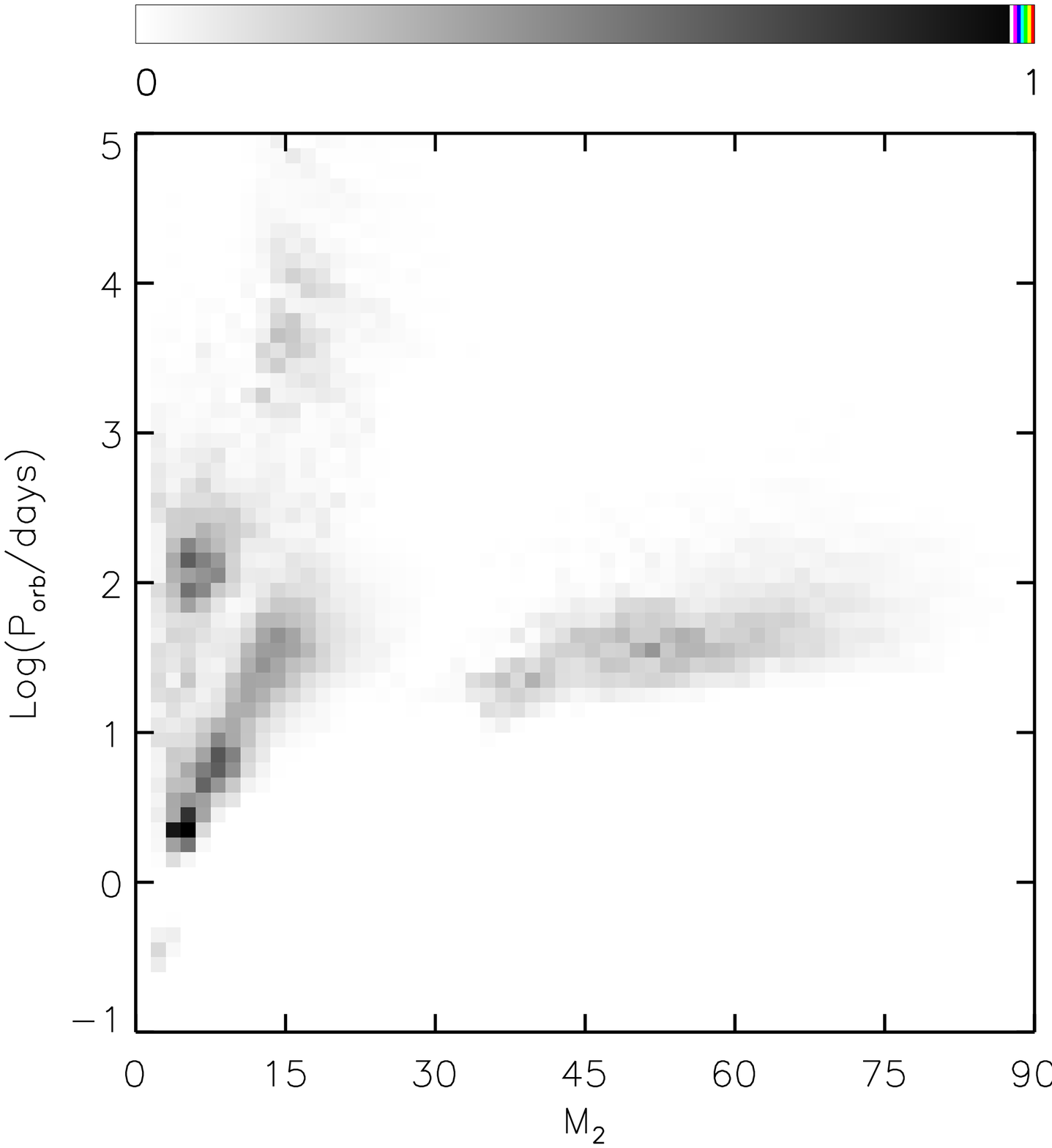}
\caption{Current orbital period $P_{\rm orb}-L_{\rm X}$ (left)
and $P_{\rm orb}-M_2$ (right) distributions in model M1.}
  \label{Fig. 1a}
\end{figure*}

The key parameters we vary include: the binary fraction $f$, the super-Eddington 
factor $\eta_{\rm Edd}$, the bolometric correction factor $\eta_{\rm bol}$, the mass ratio, 
the IMF, the natal kick distribution, the wind mass loss rates and the value of $\gamma$. Some 
parameters affect only the normalization, such as $f$ and $\eta_{\rm bol}$; some affect 
only the shape, for example, $\eta_{\rm Edd}$; while others affect both. We perform a suite 
of EPS models and find that the best-fit model in the $\gamma$ algorithm can be achieved when parameters are 
adopted as follows: SFH=50 Myr, $\alpha=0$, $\eta_{\rm Edd, BH}$=20, $f=0.5$, 
$\sigma_{\rm kick} = 110 \,\rm km\,s^{-1}$, $\eta_{\rm bol, BH}$=0.2,  
$\eta_{\rm bol, NS}$=0.1, Salpeter IMF and $\gamma=1.0$ (i.e., model M1). We also 
examine other values of $\gamma$ , and find that the results are not better than 
in the case of $\gamma=1.0$ (especially when $\gamma \gtrsim\,1.5$). In order to 
show the dependences of the XLF on the parameters, we also change these parameters 
one by one. The details are listed in Table~\ref{tab:m7}.

Figure~2 clearly shows that the parameters act in different ways. Several parameters have 
only minor effects, i.e., the wind mass loss rate (model M2) and the initial mass ratio distribution 
of the secondary star (model M5). Some (e.g. models M3 and M6) mainly increase the number of HMXB populations. An increase of the binary fraction (model M3) gives more XRBs, hence 
an overall shift of the XLF. A flatter IMF (model M6) reflects more massive stars, hence more 
compact objects that may result in XRBs. An increase of the dispersion velocity 
$\sigma_{\rm kick}$ (models M7 and M8) means 
that the natal kicks of higher magnitude are chosen more frequently from the Maxwellian 
distribution, hence more disruptions of binaries during the SN explosions. This decreases 
the number of potential HMXBs, and meanwhile changes the shape of the XLF. We note 
the large uncertainties in $\sigma_{\rm kick}$, $f$, and $\eta_{\rm bol}$  make it difficult to tightly constrain the value of $\gamma$.
The apparent luminosity `knee' of XLFs is weakened if 
we restrict the super-Eddington factor to 20 (compare model M4 with others), implying 
that the maximum super-Eddington luminosity allowed is likely $\sim 20$ in the case 
of $\gamma$ algorithm. To sum up, in the framework of the $\gamma$ algorithm, 
the observed XLF can also be reconstructed within the reasonable range of the 
parameters adopted.


In order to explore the nature of HMXBs in the case of $\gamma$ algorithm, 
we also examine the detailed observational properties (i.e., orbital period, the 
current mass $M_2$ of the donor star, etc.) of the simulated HMXBs, and compare 
them with those in \citet[][i.e, $\alpha_{\rm CE}$ formalism]{zuo14a} and observations. 
Shown in Figure~3 are the detailed components of the simulated XLF (left) and 
the accretion modes in XRBs (right) and in Figure~4 are the 
$P_{\rm orb}-L_{\rm X}$ (left) and $P_{\rm orb}-M_2$ (right) distributions in 
model M1. It is clear that under the $\gamma$ algorithm BH-He XRBs dominate 
in the low luminosity range (i.e., $L_{\rm X}<\sim10^{37}\rm\,erg\,s^{-1}$) of 
the XLF while this is not the case in the $\alpha_{\rm CE}$ formalism, 
where BH-MS XRBs instead dominate \citep{zuo14a}. Unfortunately, 
due to the limited instrument capabilities available, most of 
the extragalactic X-ray sources remain unresolved. 
We still do not clearly know their nature (for example, the spectral type of the 
donor star and the type of the compact star), especially the sources in low luminosities. We 
suggest further check with higher-precision observations is still needed in the future.
The orbital period distribution is also distinct from that in \citet{zuo14a}, with 
a much larger population of relatively short period (less than several tens of 
days) systems . This is more clearly revealed in Figure~5 for the normalized orbital period $P_{\rm orb}$ 
distribution in models A05A05 (left) and M1 (right). We can see that short 
period HMXB population keeps growing under the $\gamma$ algorithm, 
while most HMXBs under the $\alpha_{\rm CE}$ formalism are produced within 
the first 20 Myrs. These distinct observational properties of HMXBs, as well as different 
period distributions may provide further clues to discriminate between the two models.

\begin{figure*}
  \centering
 \includegraphics[width=5in]{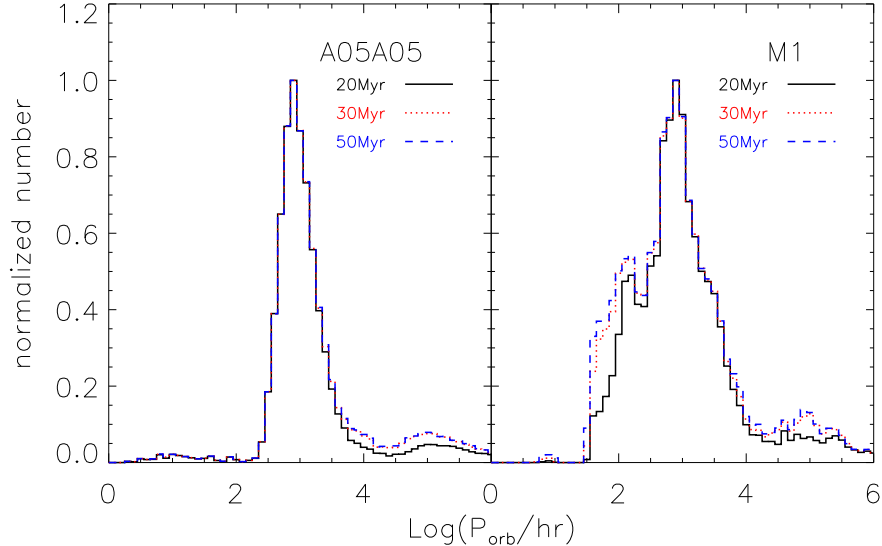}
\caption{Orbital period $P_{\rm orb}$ distribution in models A05A05 (left)
and M1 (right), respectively. The number is normalized for comparison. The solid, dotted, and
dashed lines represent SFH of 20, 30 and 50 Myr, respectively.}
  \label{Fig. 1a}
\end{figure*}

\begin{figure*}
  \centering
   \includegraphics[width=5in]{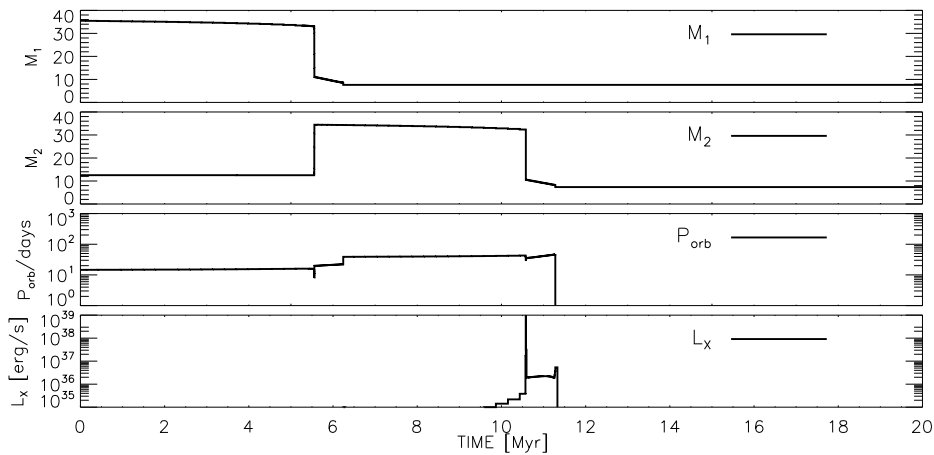}
\caption{Evolution of $M_1$, $M_2$, $P_{\rm orb}$, and $L_{\rm X}$ for
an example of short orbital period BH-He HMXBs under the $\gamma$ algorithm (model M1).}
  \label{Fig. 1a}
\end{figure*}

We note the discrepancy in the BH-He HMXB population between models is solely a result of different 
treatments on CE, in which the $\gamma$ algorithm predicts a survival, while the 
$\alpha_{\rm CE}$ formalism predicts, a merger instead. The progenitors of BH-He 
HMXBs always have the following features. First, the primary stars are very massive, 
$\sim 30-80\,M_{\odot}$, so they can form BHs in a mild (with low/no kicks) way, 
which will not disrupt the system. Second, the companion stars are relatively less massive, i.e., $\sim 10-35\,M_{\odot}$. 
The orbits of the binaries are not too wide, of the order of tens to hundreds of $R_{\odot}$. These conditions guarantee that the primary can overfill its RL rapidly and  transfer mass in a dynamically stable way (because the mass ratio is not too extreme\footnote{If the mass ratio is extreme, a CE is triggered, followed by a second CE between the newly formed BH 
and an expanding giant, instead resulting in much tighter BH-HeMS XRBs instead. However, their 
population is relatively minor in this case.}). After that, the primary evolves to a 
BH, and the rejuvenated secondary star expands and fills its RL. However, due to the large mass ratio, the mass transfer this time is unstable, and a CE is triggered. The $\alpha_{\rm CE}$ formalism always leads to binary mergers 
due to the huge amount of binding energy in the giant envelope. In the case of 
$\gamma$ algorithm, the orbital evolution is determined by the 
mass ratio $q=M_{\rm donor}/M_{\rm accretor}$ and the core mass fraction $\mu=M_{\rm c}/M_{\rm donor}$. 
From Eq.~6 in \citet{nt05}, it is easy to deduce that the binary orbit not only shrinks (but still different from 
that in the $\alpha_{\rm CE}$ formalism), but also expands (see also Figure~3 therein). This expansion of the orbit not only avoids binary mergers, but also delays the XRB formation significantly. This is also why, under the same assumptions the $\gamma$ algorithm can produce more HMXBs than the $\alpha_{\rm CE}$ formalism and the HMXBs can keep 
emerging after 20 Myr in the case of the $\gamma$ algorithm.

To illustrate the formation and evolution of a typical BH-He HMXB, 
we present one example evolutionary sequence for $M_1$, $M_2$, $P_{\rm orb}$, and 
$L_{\rm X}$ under the $\gamma$ algorithm in Figure~6. We 
consider a primordial binary system in a $\sim 91.44\,R_{\odot}$ circular orbit. 
The initial stellar masses are 35.493 and 12.532 $\,M_{\odot}$ for the primary and 
secondary, respectively. The primary evolves first, and fills its RL on the HG (at  
5.5483 Myr). The mass transfer proceeds rapidly as it evolves across the HG until
the end of CHeB, at which point (5.5598$\,$Myr) it becomes an 11.069$\,M_{\odot}$ 
HeMS star with a 34.451$\,M_{\odot}$ (rejuvenated) MS star in a $109.626\,R_{\odot}$ orbit. 
Shortly after that, the naked helium star evolves across the HeHG and collapses 
at 6.2418 Myr, leaving a  7.617$\,M_{\odot}$ BH with an MS companion in a 167.93
$\,R_{\odot}$ orbit. Subsequently, the MS star evolves to expand and fills its RL on the HG, 
and then the binary enters into the CE stage (10.5754 Myr). 
At this time, the mass ratio is $q\sim4.3$ and the core mass fraction $\mu\sim0.3$, and the 
orbit shrinks slightly from 134.38$\,R_{\odot}$ to 118.55$\,R_{\odot}$, as calculated from 
Eq.~6 in \citet{nt05}. At the end of the CE, the envelope of the giant star is expelled, leaving 
a 10.58$\,M_{\odot}$ HeMS star. The stellar wind from the HeMS star is then accreted by 
the BH, resulting in a BH-HeMS XRB. At last, the HeMS evolves to explode 
as an SN (11.28 Myr), which results in a 7.348$\,M_{\odot}$ BH and disrupts the binary system.

Our findings are generally consistent with other previous studies concerning
the CE evolution. For example, in the case of the $\alpha_{\rm CE}$ formalism, 
a high value of $\alpha_{\rm CE}$ is required to account for the observed WDMS 
PCEBs \citep[$\alpha_{\rm CE}\gtrsim 0.1$,][]{dkw10}, the shape of the delay-time 
distribution and the birth rate of SNe Ia for the double-degenerate systems \citep{my12}, 
and the displacements of HMXBs \citep[$\alpha_{\rm CE}\sim 0.8-1.0$,][]{zuo14b}, while  
a lower value of $\alpha_{\rm CE}$ may be excluded 
\citep{my12,zuo14b}. An exception is from \citet{f13} where a low value of 
$\alpha_{\rm CE} \sim 0.1$ is preferred, most likely due to the oversimplified treatments 
for the binding energy parameter, where $\lambda$ is adopted as one overall, while 
this is not the case for massive stars \citep[$\lambda \sim 0.1$,][]{xu10}.
It is interesting to note that to create double WDs, 
the standard $\alpha_{\rm CE}$ formalism is also possible if the first mass transfer between an
RG and an MS star can be stable and non-conservative. This leads to a modest widening of the orbit, 
with an effect similar to the $\gamma$ algorithm \citep{woods12}.
In the framework of the $\alpha_{\rm CE}$ formalism, 
our simulations are also comparable to previous studies concerning HMXB populations 
\citep{pv96,tts98,Linden10}. The major formation pathways of HMXBs in \citet{zuo14a} are consistent 
with the results obtained by \citet{Linden10}. The predicted observational 
properties of HMXBs (such as the orbital period distributions) are also similar. 
The number of HMXBs is also found to be not very sensitive to the value of $\alpha_{\rm CE}$. 
However, it seems that neither the $\alpha_{\rm CE}$ formalism nor the $\gamma$ algorithm can 
account for all the specific classes of observed PCEBs \citep{meng11,my12}. Moreover, even within the framework 
of the $\alpha_{\rm CE}$ formalism, different studies often give controversial results on the possible 
range of $\alpha_{\rm CE}$ and its dependence on other parameters 
\citep[see][also Ivanova et al. 2013 and references therein]{zgn00,marco11,davis12,tn13}. 
Our work suggests that in the case of HMXBs, both the $\alpha_{\rm CE}$ formalism and 
the $\gamma$ algorithm are possible to reproduce the observed XLF. In the framework of 
$\alpha_{\rm CE}$ formalism, a high value of  $\alpha_{\rm CE}$ is needed, although 
a constant value is not required. We also show the distinct observational properties, such as 
the period distribution of HMXBs, that may serve as possible keys to understanding the CE evolution 
and to discriminate between different CE models.

\section{SUMMARY}

We have used an EPS code to model the XLF of HMXBs with a range of theoretical models 
describing the CE phase. Our study shows that the observed XLF can be reproduced 
quite closely under both CE mechanisms. Provided that the same parameter combination 
is chosen, the $\gamma$ algorithm seems to produce more HMXBs 
than the $\alpha_{\rm CE}$ formalism, by a factor of up to $\sim 10$. Additionally, in the 
framework of the $\alpha_{\rm CE}$ formalism, a high value of $\alpha_{\rm CE}$ around 
$\sim 0.5-1.0$ better fits the observed XLF. We present the detailed properties of HMXB 
populations under the $\gamma$ algorithm, and find that the simulated HMXBs have a much 
larger population of short period (less than about several tens of days) BH-He systems than 
in the $\alpha_{\rm CE}$ formalism, which may serve as clues to discriminate between the 
two kinds of models. Our work motivates further high-resolution X-ray and optical observations 
of HMXB populations in nearby star-forming galaxies.

\acknowledgments  
We thank the anonymous referee for helpful suggestions that enabled us 
to improve the manuscript. This work was supported by the National 
Natural Science Foundation (under grant numbers 11103014, 11003005, 
11133001, 11333004, and 10873008), the Research Fund for the Doctoral 
Program of Higher Education of China (under grant number 20110201120034), 
the National Basic Research Program of China (973 Program 2009CB824800), 
the Strategic Priority Research Program of CAS under grant No. XDB09010200, 
and the Fundamental Research Funds for the Central Universities.

\end{document}